\documentclass[sigconf, authorversion]{acmart}

\usepackage{subcaption}

\renewcommand\footnotetextcopyrightpermission[1]{} 

\begin{document}
\title{Robust Aggregation for Adaptive Privacy Preserving Federated Learning in Healthcare}

\author{
Matei Grama, 
Maria Musat, 
Luis Mu\~noz-Gonz\'alez, 
Jonathan Passerat-Palmbach, 
Daniel Rueckert,
Amir Alansary}
\email{a.alansary14@imperial.ac.uk}
\affiliation{Imperial College London, UK}

\renewcommand{\shortauthors}{Matei Grama, et al.}




\begin{abstract}

Federated learning (FL) has enabled training models collaboratively from multiple data owning parties without sharing their data. Given the privacy regulations of patient's healthcare data, learning-based systems in healthcare can greatly benefit from privacy-preserving FL approaches.
However, typical model aggregation methods in FL are sensitive to local model updates, which may lead to failure in learning a robust and accurate global model.
In this work, we implement and evaluate different robust aggregation methods in FL applied to healthcare data. 
Furthermore, we show that such methods can detect and discard faulty or malicious local clients during training.
We run two sets of experiments using two real-world healthcare datasets for training medical diagnosis classification tasks. Each dataset is used to simulate the performance of three different robust FL aggregation strategies when facing different poisoning attacks. 
The results show that privacy preserving methods can be successfully applied alongside Byzantine-robust aggregation techniques. We observed in particular how using differential privacy (DP) did not significantly impact the final learning convergence of the different aggregation strategies. 

\end{abstract}



\maketitle

\fancyfoot{}
\thispagestyle{empty}
\pagestyle{plain} 

\section{Introduction}
During the last years the amount of digital healthcare data has grown significantly. At the same time, recent advances of deep learning (DL) have proven to achieve state-of-the-art results in several medical tasks \cite{shickel2017deep, kamnitsas2017efficient}.
DL approaches usually require a huge amount of data to train accurate models.
However, regulatory bodies have passed new laws to control sharing data while preserving user security and privacy, e.g. the general data protection regulation (GDPR) and the health insurance portability and accountability act of (HIPAA). 
Data from multiple institutions with different patient's features including image scans, medical records and academic publications, are essential to develop an intelligent and generalized DL healthcare system. 
This is challenging because of the sensitivity of sharing private patient's data under data protection regulations and laws. 
Federated learning (FL) allows DL models to train collaboratively on data from multiple institutions by sharing the learned model, while keeping local training data private \cite{mcmahan2017communication}.
However, standard distributed FL training approaches may raise a number of concerns regarding the privacy and robustness of the pipeline \cite{blanchard2017machine, he2020secure}. 
Moreover, biased local datasets, faulty clients and poisoning attacks can degrade the performance of the global aggregated model.
Very few works in the literature have evaluated the combination of privacy and robustness in FL \cite{he2020secure}.

In this paper, we study the application of robust model aggregation for FL on healthcare data. We investigate the impact of data privacy approaches such as $k$-anonymity and differential privacy (DP) over different learning tasks using tabular healthcare data. 
For this, we have modeled two poisoning attack strategies and their potential targets in diagnosis classification tasks. 
Our results are based on two datasets upon which we trained federated diagnosis classifiers: the Pima Indians diabetes dataset \cite{smith1988using} and the Cleveland heart disease Dataset \cite{detrano1989international}.

Our main contributions are: 
(1) to the best of our knowledge, this is the first work that implements and evaluates robust aggregation methods within a privacy preserving FL system applied to healthcare; 
(2) we present a survey of the state-of-the-art methods for robust aggregation in Section \ref{sec:background}; 
(3) we run an extensive set of experiments to evaluate the impact of data privacy engineering (Differential Privacy, $k$-anonymity) on the accuracy of models constructed via robust aggregation in Section \ref{sec_experiments};
(4) as part of that study, we also make available a public and unified implementation framework to allow for further experiments.



\section{Background}
\label{sec:background}

We consider the distributed FL setup \cite{mcmahan2017communication} based on a global server and $n$ clients (hospitals). 
Each client $i$ has its own private part of the training data $d_i$, which is trained locally and shared with the global server $g$. 
Whereas, $g$ aggregates these local models and update the global model, $\theta_g$, that are sent back to the clients.
This process is repeated until the training converges to a final $\theta_g$. 
In this setup, the local data that are used to train the models are not shared externally.

\subsection{Standard Aggregation}
DL models rely on stochastic gradient descent (SGD) methods for training, where the gradients are computed on a mini-batch sampled from a local dataset, and back-propagated through the model \cite{lecun2015deep}.
Shokri and Shmatikov \cite{shokri2015privacy} proposed a federated learning method based on sampling a fraction of the gradients for every client, known as FedSGD. 
The gradients are then sent to and averaged by a global server proportionally to the number of the samples on each client.

\textbf{Federated Averaging (FedAvg):}
McMahan et al. \cite{mcmahan2017communication} proposed a more efficient and generalized version of FedSGD by allowing the clients to train their local models on multiple batches of the local data. Then, the parameters of the model are shared with the global server instead of the gradients as in FedSGD. 
Hence, FedAvg can be defined as the weighted average of the local updated model parameters $\theta_i$ sent by the $n$ clients, $\theta_g=\sum_{i=1}^{n} \frac{d_{i}}{d} \theta_{i}$.
Where $d_i$ denotes the size of partitioned data used by the $i$-th client, and $d$ is the total data size used for training.
The previous equation introduces some challenges to the training process when learning from adversarial clients or heterogeneous local subsets of the data. 

\subsection{Adversarial Clients and Data Poisoning}
\label{subsec_faulty}
The main purpose of distributed FL is to learn a global model from different honest clients. 
However, in a real world scenario, medical data may contain biases, such as wrongly annotated labels or missing data and patients for instance. 
Such faulty clients sabotage the learning process by sending bad local models to the global server. 
Standard aggregation methods such as FedAvg are vulnerable to these bad local models.
In this work, similar to \cite{munoz2019byzantine}, we highlight two main strategies for bad clients, and assess their impact on the performance of the FL pipeline. 
First, \textbf{malicious clients} change the output labels of local training datasets before the training process starts. This is known as a \textit{label flipping attack}. 
Second, \textbf{Byzantine or faulty clients} send noisy parameters with every federated round, which can be classified as \textit{targeted attacks}.

\subsection{Robust Aggregation}
\label{subsec_robust}
Several methods have been published to mitigate the aforementioned limitations of FedAvg against malicious data and noisy adversaries.
The majority of these methods consider in their calculations the distribution of the local models.
In this work, we highlight the following three methods.

\subsubsection{Coordinate-wise Median (COMED):} 
Yin et al. \cite{yin2018byzantine} presented a median aggregation rule, where the global server sorts all the parameters $\theta_i$ from the $n$ clients and take the median as the selected parameter. COMED makes use of the properties of the median as a robust statistic that usually yield good performance without largely being affected by outliers. However, it assumes that each client shares the same size of data during training, which may cause the aggregation rules to be inefficient when the clients share partitions of the training data with different sizes. Moreover, the complexity of calculating the median scales up with the number of clients and model's parameters.

\subsubsection{Multi-Krum (MKRUM):}
Blanchard et al. \cite{blanchard2017machine} proposed a robust aggregation method based on selecting a single or multiple $m$ local models that are most similar to each other, known as multi-Krum (MKRUM).
Although, the elimination of some models may impact the performance of the training process, it constrains the global server to only aggregate similar models. 
The Krum function assigns a score for each client $i$ based on the summation of the Euclidean distances between $i$ and the rest of clients, $s_i=\sum^{n}_{j=0} \left\|V_{i}-V_{j}\right\|^{2}$ for all $i\neq j$.
Finally, MKRUM selects and aggregates the $m$ local models
with the smallest sum of distance as the global model, then a standard FedAvg is applied only to the selected models. 
Similar to COMED, the complexity of MKRUM's computations can scale up with the number of clients and model's parameters.

\subsubsection{Adaptive Federated Averaging (AFA):}
Mu{\~n}oz-Gonz{\'a}lez et al. \cite{munoz2019byzantine} proposed an adaptive aggregation method that models the client's behavior with a hidden Markov model (HMM), which allows to block clients that systematically send bad model updates, reducing the computational and the communication burden in the presence of malicious and faulty clients. 
AFA's robust aggregation algorithm relies on computing the similarities between the model updates provided by the clients, $\theta_i$, and the aggregated model parameters $\theta_g$. 
Then, assuming that the number of benign clients is more than half the total number of clients, at each training round, a cluster of clients with a \emph{similar} similarity score with respect to the global model are selected, discarding the updates sent by the others. 
The algorithm includes hyper-parameters to allow for more or less flexibility to accept or reject local updates, depending on the scenario and the heterogeneity of the datasets. 
Given the set of benign clients selected at each iteration, $K^{g}$, the parameters of the global model are updated as:
\begin{equation}
\theta_g \leftarrow \sum_{k \subset K^{g}} \frac{p_k d_k}{\sum_{k \subset K^{g}} p_{k} d_{k} \theta_{k}}.
\end{equation} 
\noindent Where $d_k$ is the number of training points provided by each selected client, and $p_k$ is the probability of client $k$ providing good model updates. 
This probability is updated with the HMM that models the client's behavior by taking into account the good and bad model updates provided at each training round. 

\subsection{Input Data Privacy}
\label{subsec_privacy}
In the standard FL pipeline, private information can be inferred from the shared models, which can be traced back to its source in the resulting trained model \cite{nasr2018comprehensive}.
For example, Shokri et al. \cite{shokri2017membership} have shown that the participation of individual patients can be identified from the trained models on private medical data from a particular hospital.
Previous works have been proposed to control privacy exposure and provide reasonable data privacy guarantees.
Here, we highlight two main privacy preserving methods that are not computationally expensive and commonly used in the literature, namely \textbf{differential privacy (DP)} \cite{dwork2008differential} and \textbf{k-anonymity} \cite{sweeney2002k}. 






\section{Data Privacy Combined with Robust Aggregation FL}
\label{sec_method}
Data privacy methods (Section \ref{subsec_privacy}) are designed to protect the privacy of user information by injecting a certain amount of noise. 
This can mitigate the performance of robust aggregation methods (Section \ref{subsec_robust}) that rely on distance measures to identify any potential adversarial clients.

\subsection{Differential Privacy (DP)}
Similar to \cite{li2019privacy}, we apply $\epsilon$-DP on the client-side in the robust FL setup, by selecting and sharing distorted sets of $\theta_i$. 
In DP, an algorithm is considered to be differentially private, if and only if a single data entry does not significantly affect the algorithm's output.
This is usually done by applying random noise at different levels of the training algorithm in order to increase data privacy.
A set of four parameters have to be tuned in order to adjust the random noise levels \cite{shokri2015privacy}. 
These parameters in turn control the privacy level according to the privacy budget, and the overall performance of the model.
For example, the gradient clipping threshold $\gamma$ and the sensitivity of the gradient $s$ play a role in computing the added noise. 
$Q$ controls the release proportion of the trained parameters
Three $\epsilon$ values are used to tune the Laplacian noise levels according to the privacy budget: $\epsilon_1$ adjusts the noise added for randomly selecting parameters to release and $\epsilon_3$ adjusts the noise added to the released parameters. 
Finally, $\epsilon_2$ controls the noisy threshold used for selecting the subset of parameters to be released, which is computed based on $\epsilon_1$, $s$ and $Q$.

\subsection{k-anonymity}
We characterize a syntactic approach \cite{choudhury2020anonymizing} for data privacy that relies on generalization of the identifying information from relational datasets. 
It is designed to minimize the information loss and preserve analytical values, which is inevitably caused by enforcing data protection through anonymization.
The $k$-anonymity approach anonymizes quasi-identifying attributes that could allow a malicious user to identify or disclose sensitive information about an individual.
The $k$-anonymity property is then enforced for each of the client's datasets after selecting the quasi-identifiers, where each quasi-identifier tuple occurs at least $k$ entries.
This mechanism guarantees that a single entry cannot be differentiated from at least $k-1$ others. 
Then, syntactic mappings are extracted corresponding to the enforced dataset transformation, and collected from all the clients.
The same mappings are used to anonymize the test dataset owned by the aggregator.
As a result of this hierarchical mapping of the relational attributes, $k$-anonymity can enforce a mean of data generalization that can help a more robust aggregation. 
The next section shows how both the syntactic approach and the client-level DP have been implemented alongside Byzantine-robust aggregation schemes in order to evaluate the trade-offs of those techniques, in the context of medical applications.

\section{Experiments and Results}
\label{sec_experiments}

The performance of the different robust aggregation methods discussed in \ref{subsec_robust}, FedAvg (FA), MKRUM, COMED and AFA, are evaluated using two experiments on two different healthcare datasets. The implementation of our code will be publicly available on GitHub, and shared in the camera ready version of the paper.

\subsection{Experiment-I:} 
\label{subsec_experiment1}
The main task of this experiment is to learn federated models that can predict the onset of the diabetes mellitus condition.

\textbf{Dataset:}
We use the \emph{Pima Indians Diabetes} dataset from the National Institute of Diabetes and Digestive and Kidney Diseases \cite{smith1988using}. 
Data entries consist of medical measurements that correspond to $768$ female patients older than $21$ years old. 
These measurements include diastolic blood pressure, body mass index, and plasma glucose concentration.
Data is split into $614$ samples for training and $154$ for testing. 
The training data are shared among $10$ federated clients, where three of them own $39$ samples, three own $59$ and four own $80$.
Client $1$ is used as faulty, and clients $(2,4)$ are malicious.

\textbf{Implementation:}
All the clients share the same neural network (NN) model, which consists of $3$ fully-convolutional (FC) layers of size $[200, 200, 2]$ neurons and combined with ReLU activation functions. 
Local training epochs are fixed to $5$ iterations, and $50$ federated rounds. 
We use a learning rate of $1e-5$ with an Adam optimizer, batch size of $10$, and cross entropy as a loss function.
The client-level of the DP module is set to release $10\%$ of the clients' trained parameters, after adding the Laplacian noise with $\epsilon_1=\epsilon_3=1e-4$.
The values of the DP parameters are set similar to \cite{li2019privacy}.
For the $k$-anonymity experiments, we set $k$ arbitrarily to $4$ with respect to the age and pregnancies attributes.
The value of $k$ is usually selected according to the data protection regulations that address the training data usage. 

\begin{figure*}[ht]
    \begin{subfigure}[tb]{0.49\linewidth}
        \centering
        \includegraphics[scale=0.34, keepaspectratio, trim=6 8 660 20, clip]{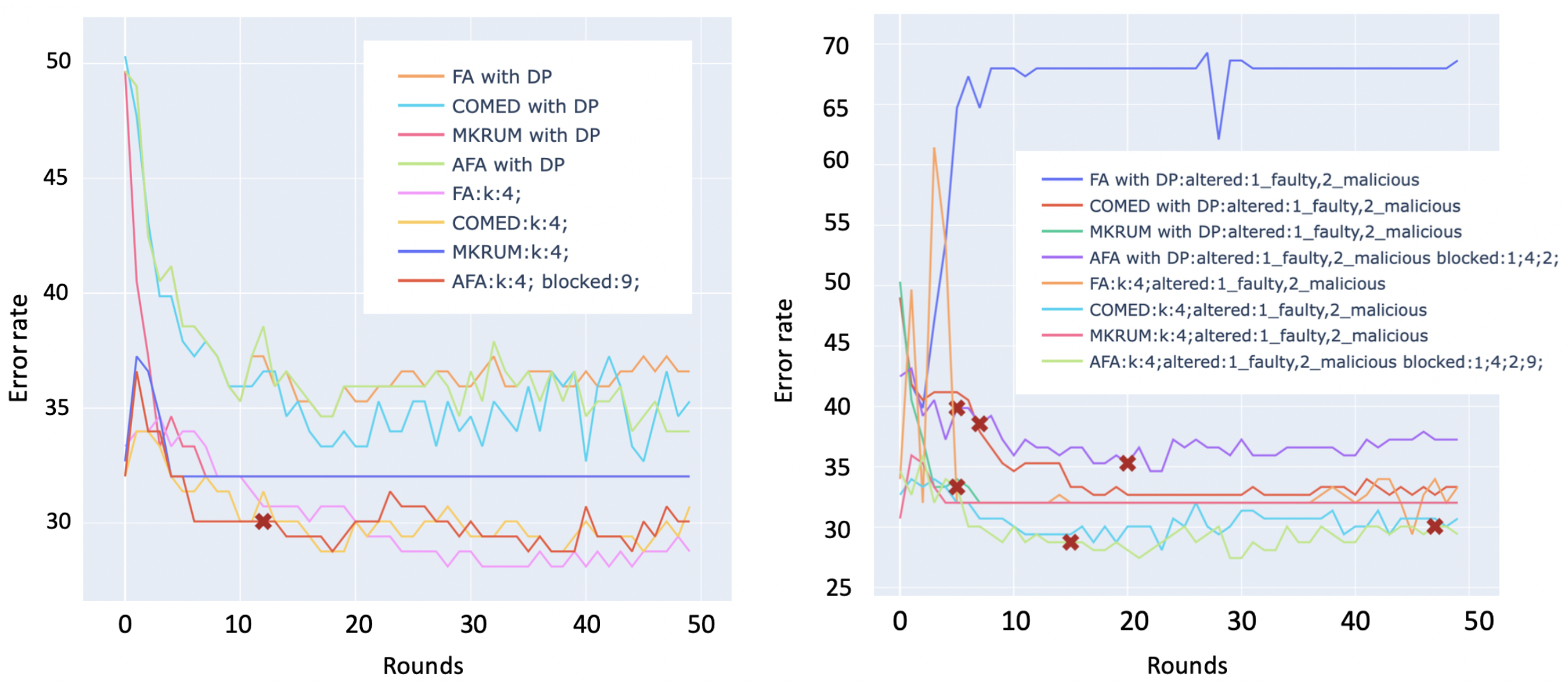}
        \caption{Experiment-I: without bad clients.}
        \label{fig:diabetes_1}
    \end{subfigure}
    \begin{subfigure}[tb]{0.49\linewidth}
        \centering
        \includegraphics[scale=0.34, keepaspectratio, trim=604 8 20 20, clip]{./pics/diabetes.png}
        \caption{Experiment-I: client 1 faulty and client (2,4) malicious.}
        \label{fig:diabetes_2}
    \end{subfigure}
        
    \begin{subfigure}[tb]{0.49\linewidth}
    \centering
    \includegraphics[scale=0.34, keepaspectratio, trim=6 25 655 20, clip]{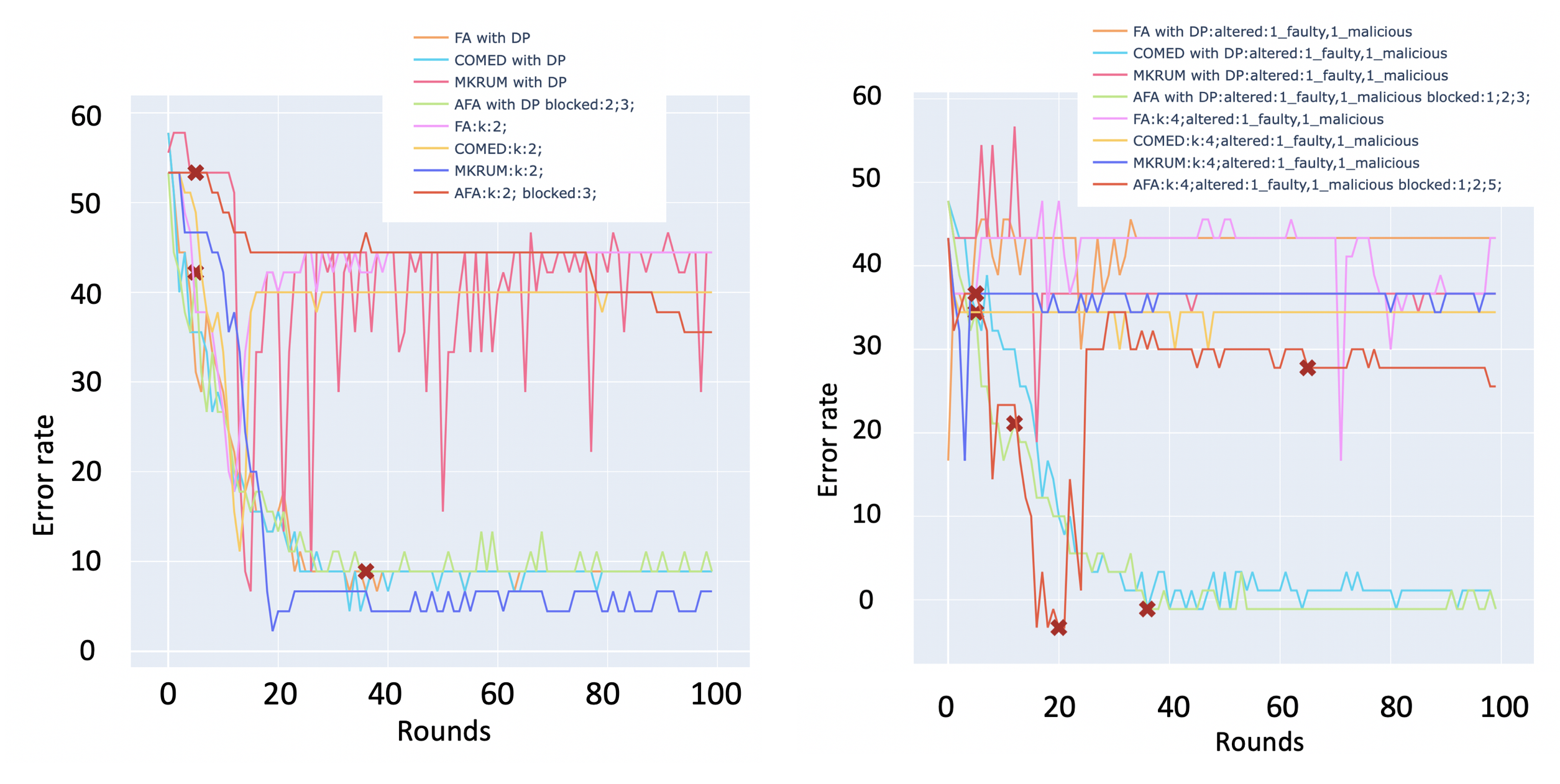} 
    \caption{Experiment-II: without bad clients.}
    \label{fig:heart_1}
\end{subfigure} 
\begin{subfigure}[tb]{0.49\linewidth}
    \centering
    \includegraphics[scale=0.34, keepaspectratio, trim=655 25 20 20, clip]{./pics/heart_all.png} 
    \caption{Experiment-II: client 1 faulty and client 2 malicious.}
    \label{fig:heart_2}
\end{subfigure} 
    \caption{Performance of privacy preserving robust aggregation. \textit{The red cross indicates the detection of a suspected bad client.}}
\end{figure*}

\textbf{Results:}
Figure \ref{fig:diabetes_1} demonstrates the performance of different aggregation methods with no bad clients.
The results from the DP experiments show that AFA and COMED can be more sensitive than FA and MKRUM to the DP noise added to the local model parameters. 
While the results from the syntactic approach with $k=4$ show a similar performance for all the models with a slightly lower error rate for AFA and a higher one for MKRUM. 
Figure \ref{fig:diabetes_2} demonstrates the performance of different aggregation methods with one faulty and one malicious clients.
The results from the DP experiments show that model trained using FA diverges compared to the rest of the robust aggregation methods, while MKRUM performs the best.
The results from the syntactic approach are similar to the experiment with no faulty clients, despite the unstable performance of FA during the first rounds. 
AFA detects and blocks the malignant clients when used with DP and $k$-anonymity. 
However, it may also block benign clients if its parameters are not fine-tuned. 
We also noticed that the final learning convergence is not greatly affected, as blocking benign clients usually happens later in the training.
Moreover, the results show that the $k$-anonymity approach improves the robustness of standard aggregation methods.



\subsection{Experiment-II:} 
\label{subsec_experiment2}
The main task in this experiment is to learn federated models that can predict heart failure.

\textbf{Dataset:}
We use the Cleveland heart disease database \cite{detrano1989international}, from which we select $297$ data samples with $14$ attributes for each. 
The data is split to $207$ entries for training and $46$ entries for testing.
The data attributes include personal and medical measurements such as: age, sex, fasting blood sugar, and maximum heart rate achieved.
Given the small size of the data set, we equally split the data between $5$ federated clients. 
One malicious and one faulty clients are simulated for the experiment with bad clients.

\textbf{Implementation:}
All the clients share the same NN model, which consists of $3$ FC layers of size $[32, 16, 2]$ neurons and combined with ReLU activation functions.
Each client share the same number of $10$ training epochs, and $100$ federated rounds.
We use a learning rate of $1e-4$ with an Adam optimizer, batch size of $5$, and a cross entropy loss function.
Data privacy parameters are configured similar to Experiment-I in Section \ref{subsec_experiment1}.


\textbf{Results:}
Figure \ref{fig:heart_1} demonstrates the performance of different aggregation methods with no bad clients. 
The DP experiments show that MKRUM fails to learn an accurate model, while the rest of the methods perform similarly with error rate around $5-9\%$.
Using the syntactic approach with $k=4$, MKRUM achieves the best performance with $4-6\%$ error rate, while the rest of the methods fail to learn an accurate model.
Figure \ref{fig:heart_2} demonstrates the performance of different aggregation methods with one malicious and one faulty client.
DP experiments show that FA and MKRUM fail to learn a global accurate model, while AFA and COMED performs the best with error rate close to $9\%$. 
AFA was able to detect and block three suspected clients, from which were two faulty and malicious clients. 
The blocked benign client did not affect the convergence of the final model.
The results from the syntactic approach show that all methods fail to converge to an accurate model.
Although AFA managed to detect and block the bad clients in federated round $5$ and $21$, the error started to diverge in the later rounds.
In contrary to the results from Experiment-I, $k$-anonymity did not improve the robustness of the aggregation methods. 
This decline of performance in the syntactic approach could result from the smaller size of the heart dataset compared to the diabetes dataset. 
\section{Conclusion}
In this paper, we study the application of robust aggregation for adaptive federated learning (FL) to healthcare data.
We evaluate the performance of different robust aggregation methods with differential privacy (DP) and $k$-anonymity on top of the standard FL setup.
We use two healthcare dataset for predicting diabetes and heart failure from medical records.
Our experiments show that the syntactic approach using $k=4$ can help a more robust aggregation, if it is applied to a dataset with a sufficient number of samples. 
Moreover, in scenarios where faulty or malicious clients exist, adaptive federated averaging (AFA) consistently outperforms other aggregation methods, while still being able to detect and block bad clients.
In the future, we will explore the generalization of AFA's specific hyper-parameters that can translate into increased performance for the learning process in robust and privacy preserving settings.
We will also explore the impact of data size and the number of faulty and malicious clients on the performance of different aggregation methods.

\vfill
\newpage

\balance
\bibliographystyle{ACM-Reference-Format}
\bibliography{references}


\begin{thebibliography}{17}


\ifx \showCODEN    \undefined \def \showCODEN     #1{\unskip}     \fi
\ifx \showDOI      \undefined \def \showDOI       #1{#1}\fi
\ifx \showISBNx    \undefined \def \showISBNx     #1{\unskip}     \fi
\ifx \showISBNxiii \undefined \def \showISBNxiii  #1{\unskip}     \fi
\ifx \showISSN     \undefined \def \showISSN      #1{\unskip}     \fi
\ifx \showLCCN     \undefined \def \showLCCN      #1{\unskip}     \fi
\ifx \shownote     \undefined \def \shownote      #1{#1}          \fi
\ifx \showarticletitle \undefined \def \showarticletitle #1{#1}   \fi
\ifx \showURL      \undefined \def \showURL       {\relax}        \fi
\providecommand\bibfield[2]{#2}
\providecommand\bibinfo[2]{#2}
\providecommand\natexlab[1]{#1}
\providecommand\showeprint[2][]{arXiv:#2}

\bibitem[\protect\citeauthoryear{Blanchard, Guerraoui, Stainer,
  et~al\mbox{.}}{Blanchard et~al\mbox{.}}{2017}]%
        {blanchard2017machine}
\bibfield{author}{\bibinfo{person}{Peva Blanchard}, \bibinfo{person}{Rachid
  Guerraoui}, \bibinfo{person}{Julien Stainer}, {et~al\mbox{.}}}
  \bibinfo{year}{2017}\natexlab{}.
\newblock \showarticletitle{Machine learning with adversaries: Byzantine
  tolerant gradient descent}. In \bibinfo{booktitle}{\emph{Advances in Neural
  Information Processing Systems}}. \bibinfo{pages}{119--129}.
\newblock


\bibitem[\protect\citeauthoryear{Choudhury, Gkoulalas-Divanis, Salonidis,
  Sylla, Park, Hsu, and Das}{Choudhury et~al\mbox{.}}{2020}]%
        {choudhury2020anonymizing}
\bibfield{author}{\bibinfo{person}{Olivia Choudhury}, \bibinfo{person}{Aris
  Gkoulalas-Divanis}, \bibinfo{person}{Theodoros Salonidis},
  \bibinfo{person}{Issa Sylla}, \bibinfo{person}{Yoonyoung Park},
  \bibinfo{person}{Grace Hsu}, {and} \bibinfo{person}{Amar Das}.}
  \bibinfo{year}{2020}\natexlab{}.
\newblock \showarticletitle{Anonymizing Data for Privacy-Preserving Federated
  Learning}.
\newblock \bibinfo{journal}{\emph{arXiv preprint arXiv:2002.09096}}
  (\bibinfo{year}{2020}).
\newblock


\bibitem[\protect\citeauthoryear{Detrano, Janosi, Steinbrunn, Pfisterer,
  Schmid, Sandhu, Guppy, Lee, and Froelicher}{Detrano et~al\mbox{.}}{1989}]%
        {detrano1989international}
\bibfield{author}{\bibinfo{person}{Robert Detrano}, \bibinfo{person}{Andras
  Janosi}, \bibinfo{person}{Walter Steinbrunn}, \bibinfo{person}{Matthias
  Pfisterer}, \bibinfo{person}{Johann-Jakob Schmid}, \bibinfo{person}{Sarbjit
  Sandhu}, \bibinfo{person}{Kern~H Guppy}, \bibinfo{person}{Stella Lee}, {and}
  \bibinfo{person}{Victor Froelicher}.} \bibinfo{year}{1989}\natexlab{}.
\newblock \showarticletitle{International application of a new probability
  algorithm for the diagnosis of coronary artery disease}.
\newblock \bibinfo{journal}{\emph{The American journal of cardiology}}
  \bibinfo{volume}{64}, \bibinfo{number}{5} (\bibinfo{year}{1989}),
  \bibinfo{pages}{304--310}.
\newblock


\bibitem[\protect\citeauthoryear{Dwork}{Dwork}{2008}]%
        {dwork2008differential}
\bibfield{author}{\bibinfo{person}{Cynthia Dwork}.}
  \bibinfo{year}{2008}\natexlab{}.
\newblock \showarticletitle{Differential privacy: A survey of results}. In
  \bibinfo{booktitle}{\emph{International conference on theory and applications
  of models of computation}}. Springer, \bibinfo{pages}{1--19}.
\newblock


\bibitem[\protect\citeauthoryear{He, Karimireddy, and Jaggi}{He
  et~al\mbox{.}}{2020}]%
        {he2020secure}
\bibfield{author}{\bibinfo{person}{Lie He}, \bibinfo{person}{Sai~Praneeth
  Karimireddy}, {and} \bibinfo{person}{Martin Jaggi}.}
  \bibinfo{year}{2020}\natexlab{}.
\newblock \showarticletitle{Secure Byzantine-Robust Machine Learning}.
\newblock \bibinfo{journal}{\emph{arXiv preprint arXiv:2006.04747}}
  (\bibinfo{year}{2020}).
\newblock


\bibitem[\protect\citeauthoryear{Kamnitsas, Ledig, Newcombe, Simpson, Kane,
  Menon, Rueckert, and Glocker}{Kamnitsas et~al\mbox{.}}{2017}]%
        {kamnitsas2017efficient}
\bibfield{author}{\bibinfo{person}{Konstantinos Kamnitsas},
  \bibinfo{person}{Christian Ledig}, \bibinfo{person}{Virginia~FJ Newcombe},
  \bibinfo{person}{Joanna~P Simpson}, \bibinfo{person}{Andrew~D Kane},
  \bibinfo{person}{David~K Menon}, \bibinfo{person}{Daniel Rueckert}, {and}
  \bibinfo{person}{Ben Glocker}.} \bibinfo{year}{2017}\natexlab{}.
\newblock \showarticletitle{Efficient multi-scale 3D CNN with fully connected
  CRF for accurate brain lesion segmentation}.
\newblock \bibinfo{journal}{\emph{Medical image analysis}}
  \bibinfo{volume}{36} (\bibinfo{year}{2017}), \bibinfo{pages}{61--78}.
\newblock


\bibitem[\protect\citeauthoryear{LeCun, Bengio, and Hinton}{LeCun
  et~al\mbox{.}}{2015}]%
        {lecun2015deep}
\bibfield{author}{\bibinfo{person}{Yann LeCun}, \bibinfo{person}{Yoshua
  Bengio}, {and} \bibinfo{person}{Geoffrey Hinton}.}
  \bibinfo{year}{2015}\natexlab{}.
\newblock \showarticletitle{Deep learning}.
\newblock \bibinfo{journal}{\emph{nature}} \bibinfo{volume}{521},
  \bibinfo{number}{7553} (\bibinfo{year}{2015}), \bibinfo{pages}{436--444}.
\newblock


\bibitem[\protect\citeauthoryear{Li, Milletar{\`\i}, Xu, Rieke, Hancox, Zhu,
  Baust, Cheng, Ourselin, Cardoso, et~al\mbox{.}}{Li et~al\mbox{.}}{2019}]%
        {li2019privacy}
\bibfield{author}{\bibinfo{person}{Wenqi Li}, \bibinfo{person}{Fausto
  Milletar{\`\i}}, \bibinfo{person}{Daguang Xu}, \bibinfo{person}{Nicola
  Rieke}, \bibinfo{person}{Jonny Hancox}, \bibinfo{person}{Wentao Zhu},
  \bibinfo{person}{Maximilian Baust}, \bibinfo{person}{Yan Cheng},
  \bibinfo{person}{S{\'e}bastien Ourselin}, \bibinfo{person}{M~Jorge Cardoso},
  {et~al\mbox{.}}} \bibinfo{year}{2019}\natexlab{}.
\newblock \showarticletitle{Privacy-preserving federated brain tumour
  segmentation}. In \bibinfo{booktitle}{\emph{International Workshop on Machine
  Learning in Medical Imaging}}. Springer, \bibinfo{pages}{133--141}.
\newblock


\bibitem[\protect\citeauthoryear{McMahan, Moore, Ramage, Hampson, and
  y~Arcas}{McMahan et~al\mbox{.}}{2017}]%
        {mcmahan2017communication}
\bibfield{author}{\bibinfo{person}{Brendan McMahan}, \bibinfo{person}{Eider
  Moore}, \bibinfo{person}{Daniel Ramage}, \bibinfo{person}{Seth Hampson},
  {and} \bibinfo{person}{Blaise~Aguera y Arcas}.}
  \bibinfo{year}{2017}\natexlab{}.
\newblock \showarticletitle{Communication-efficient learning of deep networks
  from decentralized data}. In \bibinfo{booktitle}{\emph{Artificial
  Intelligence and Statistics}}. \bibinfo{pages}{1273--1282}.
\newblock


\bibitem[\protect\citeauthoryear{Mu{\~n}oz-Gonz{\'a}lez, Co, and
  Lupu}{Mu{\~n}oz-Gonz{\'a}lez et~al\mbox{.}}{2019}]%
        {munoz2019byzantine}
\bibfield{author}{\bibinfo{person}{Luis Mu{\~n}oz-Gonz{\'a}lez},
  \bibinfo{person}{Kenneth~T Co}, {and} \bibinfo{person}{Emil~C Lupu}.}
  \bibinfo{year}{2019}\natexlab{}.
\newblock \showarticletitle{Byzantine-robust federated machine learning through
  adaptive model averaging}.
\newblock \bibinfo{journal}{\emph{arXiv preprint arXiv:1909.05125}}
  (\bibinfo{year}{2019}).
\newblock


\bibitem[\protect\citeauthoryear{Nasr, Shokri, and Houmansadr}{Nasr
  et~al\mbox{.}}{2018}]%
        {nasr2018comprehensive}
\bibfield{author}{\bibinfo{person}{Milad Nasr}, \bibinfo{person}{Reza Shokri},
  {and} \bibinfo{person}{Amir Houmansadr}.} \bibinfo{year}{2018}\natexlab{}.
\newblock \showarticletitle{Comprehensive privacy analysis of deep learning:
  Stand-alone and federated learning under passive and active white-box
  inference attacks}.
\newblock \bibinfo{journal}{\emph{arXiv preprint arXiv:1812.00910}}
  (\bibinfo{year}{2018}).
\newblock


\bibitem[\protect\citeauthoryear{Shickel, Tighe, Bihorac, and Rashidi}{Shickel
  et~al\mbox{.}}{2017}]%
        {shickel2017deep}
\bibfield{author}{\bibinfo{person}{Benjamin Shickel},
  \bibinfo{person}{Patrick~James Tighe}, \bibinfo{person}{Azra Bihorac}, {and}
  \bibinfo{person}{Parisa Rashidi}.} \bibinfo{year}{2017}\natexlab{}.
\newblock \showarticletitle{Deep EHR: a survey of recent advances in deep
  learning techniques for electronic health record (EHR) analysis}.
\newblock \bibinfo{journal}{\emph{IEEE journal of biomedical and health
  informatics}} \bibinfo{volume}{22}, \bibinfo{number}{5}
  (\bibinfo{year}{2017}), \bibinfo{pages}{1589--1604}.
\newblock


\bibitem[\protect\citeauthoryear{Shokri and Shmatikov}{Shokri and
  Shmatikov}{2015}]%
        {shokri2015privacy}
\bibfield{author}{\bibinfo{person}{Reza Shokri} {and} \bibinfo{person}{Vitaly
  Shmatikov}.} \bibinfo{year}{2015}\natexlab{}.
\newblock \showarticletitle{Privacy-preserving deep learning}. In
  \bibinfo{booktitle}{\emph{Proceedings of the 22nd ACM SIGSAC conference on
  computer and communications security}}. \bibinfo{pages}{1310--1321}.
\newblock


\bibitem[\protect\citeauthoryear{Shokri, Stronati, Song, and Shmatikov}{Shokri
  et~al\mbox{.}}{2017}]%
        {shokri2017membership}
\bibfield{author}{\bibinfo{person}{Reza Shokri}, \bibinfo{person}{Marco
  Stronati}, \bibinfo{person}{Congzheng Song}, {and} \bibinfo{person}{Vitaly
  Shmatikov}.} \bibinfo{year}{2017}\natexlab{}.
\newblock \showarticletitle{Membership inference attacks against machine
  learning models}. In \bibinfo{booktitle}{\emph{2017 IEEE Symposium on
  Security and Privacy (SP)}}. IEEE, \bibinfo{pages}{3--18}.
\newblock


\bibitem[\protect\citeauthoryear{Smith, Everhart, Dickson, Knowler, and
  Johannes}{Smith et~al\mbox{.}}{1988}]%
        {smith1988using}
\bibfield{author}{\bibinfo{person}{Jack~W Smith}, \bibinfo{person}{JE
  Everhart}, \bibinfo{person}{WC Dickson}, \bibinfo{person}{WC Knowler}, {and}
  \bibinfo{person}{RS Johannes}.} \bibinfo{year}{1988}\natexlab{}.
\newblock \showarticletitle{Using the ADAP learning algorithm to forecast the
  onset of diabetes mellitus}. In \bibinfo{booktitle}{\emph{Proceedings of the
  Annual Symposium on Computer Application in Medical Care}}. American Medical
  Informatics Association, \bibinfo{pages}{261}.
\newblock


\bibitem[\protect\citeauthoryear{Sweeney}{Sweeney}{2002}]%
        {sweeney2002k}
\bibfield{author}{\bibinfo{person}{Latanya Sweeney}.}
  \bibinfo{year}{2002}\natexlab{}.
\newblock \showarticletitle{k-anonymity: A model for protecting privacy}.
\newblock \bibinfo{journal}{\emph{International Journal of Uncertainty,
  Fuzziness and Knowledge-Based Systems}} \bibinfo{volume}{10},
  \bibinfo{number}{05} (\bibinfo{year}{2002}), \bibinfo{pages}{557--570}.
\newblock


\bibitem[\protect\citeauthoryear{Yin, Chen, Kannan, and Bartlett}{Yin
  et~al\mbox{.}}{2018}]%
        {yin2018byzantine}
\bibfield{author}{\bibinfo{person}{Dong Yin}, \bibinfo{person}{Yudong Chen},
  \bibinfo{person}{Ramchandran Kannan}, {and} \bibinfo{person}{Peter
  Bartlett}.} \bibinfo{year}{2018}\natexlab{}.
\newblock \showarticletitle{Byzantine-Robust Distributed Learning: Towards
  Optimal Statistical Rates}. In \bibinfo{booktitle}{\emph{International
  Conference on Machine Learning}}. \bibinfo{pages}{5650--5659}.
\newblock


\end{thebibliography}

\end{document}